\documentclass[11pt,a4paper]{article}                           

\usepackage[bookmarksopen, bookmarksnumbered, bookmarksopenlevel=2]{hyperref}
\usepackage{tikz}
\usepackage{tikz-3dplot}
\usetikzlibrary{calc}
\usetikzlibrary{decorations} %
\usepackage[UKenglish]{babel}
\usepackage[toc,page]{appendix}
\usepackage{amsmath}
\usepackage{amssymb}
\usepackage{graphicx}
\usepackage{hhline}
\usepackage[bf]{caption}
\usepackage{cite}
\usepackage[vcentermath]{youngtab}
\usepackage{geometry}
\usepackage{slashed}
\usepackage{graphicx} 
\usepackage[T1]{fontenc} 
\usepackage[utf8]{inputenc}
\newcommand{\bqa}{\begin{eqnarray}}
\newcommand{\eqa}{\end{eqnarray}}

\newcommand{\be}{\begin{equation}}
\newcommand{\ee}{\end{equation}}
\newcommand{\ba}{\begin{aligned}}
\newcommand{\ea}{\end{aligned}}

\hypersetup{
    pdftitle={},
    pdfauthor={},
    pdfsubject={}
}
\numberwithin{equation}{section}
\numberwithin{table}{section}

\begin{document}

\baselineskip=14pt
\parskip 5pt plus 1pt

\vspace*{-1.5cm}
\begin{flushright}    % Publication numbers
  {\small

  }
\end{flushright}

\vspace{2cm}
\begin{center}        % Main title
  {\LARGE On TCS $G_2$ manifolds and 4D Emergent Strings }
\end{center}

\vspace{0.75cm}
\begin{center}        % Authors
Fengjun Xu
\end{center}
\vspace{0.15cm}
\begin{center}        % Institutes
%{}^a\emph{Arnold-Sommerfeld-Zentrum f\"ur Theoretische Physik\\
%             Fakult\"at f\"ur Physik, Ludwig-Maximilians-Universit\"at M\"unchen\\
%             Theresienstra\ss{}e 37, 80333 München, Germany
%             }\\[2mm]
\emph{  Institut f\"ur Theoretische Physik, Ruprecht-Karls-Universit\"at, 
\\       Philosophenweg 19, 69120,
Heidelberg, Germany\\
          }
\end{center}

\vspace{2cm}

%%%%%%%%%%%%%%%%%%%%%%%%%%%%%%%%%%%%%%%%%%%%%%%
%%%%%%%%%%%%%%%%%%%%%%%%%%%%%%%%%%%%%%%%%%%%%%%
%%%%%%%%%%%%%%%%%%%%%%%%%%%%%%%%%%%%%%%%%%%%%%%

\begin{abstract}
\noindent  In this note, we study the Swampland Distance Conjecture in TCS $G_2$ manifold compactifications of M-theory.  In particular, we are interested in testing a refined version- the Emergent String Conjecture, in settings with 4d $N=1$ supersymmetry. We find that a weakly coupled, tensionless fundamental heterotic string does emerge at the infinite distance limit characterized by shrinking the $K3$-fiber in a TCS $G_2$ manifold. Such a fundamental tensionless string leads to the parametrically leading infinite tower of asymptotically massless states, which is in line with the Emergent String Conjecture. The tensionless string, however, receives quantum corrections.  We check that these quantum corrections do modify the volume of the shrinking $K3$-fiber via string duality and hence make the string regain a non-vanishing tension at the quantum level, leading to a decompactification. Geometrically, the quantum corrections modify the metric of the classical moduli space and are expected to obstruct the infinite distance limit.  We also comment on another possible type of infinite distance limit in TCS $G_2$ compactifications, which might lead to a weakly coupled fundamental type II string theory. 

\end{abstract}

\thispagestyle{empty}
\clearpage
\tableofcontents
\thispagestyle{empty}

%%%%%%%%%%%%%%%%%%%%%%%%%%%%%%%%%%%%%%%%%%%%%%%
%%%%%%%%%%%%%%%%%%%%%%%%%%%%%%%%%%%%%%%%%%%%%%%
%%%%%%%%%%%                 %%%%%%%%%%%%%%%%%%%
%%%%%%%%%%%  DOCUMENT BODY  %%%%%%%%%%%%%%%%%%%
%%%%%%%%%%%                 %%%%%%%%%%%%%%%%%%%
%%%%%%%%%%%%%%%%%%%%%%%%%%%%%%%%%%%%%%%%%%%%%%%
%%%%%%%%%%%%%%%%%%%%%%%%%%%%%%%%%%%%%%%%%%%%%%%
%%%%%%%%%%%%%%%%%%%%%%%%%%%%%%%%%%%%%%%%%%%%%%%

\section{Introduction}

Lots of efforts have been recently devoted to the so-called swampland program, of which the main goal is to characterize a set of criteria, which effective theories must satisfy to be low-energy descriptions of a UV-complete theory of quantum gravity. The term"swampland" \cite{Vafa:2005ui}, contrasted with the "Landscape", refers to the consistent quantum field theories that cannot be consistently coupled to quantum gravity at the UV. All these criteria, though at the intermediate state as being conjectures, would be of particular interest for phenomenology, especially in cosmology and particle physics.  We refer to \cite{Palti:2019pca, Brennan:2017rbf} for recent overviews on this program. 

 Among these swampland conjectures \cite{Polchinski:2003bq, ArkaniHamed:2006dz, Ooguri:2006in, Palti:2017elp, Obied:2018sgi, Ooguri:2018wrx, Klaewer:2018yxi, Lust:2019zwm, McNamara:2019rup, Bedroya:2019snp, Palti:2020tsy}, it appears that the Swampland Distance Conjecture (SDC) \cite{Ooguri:2006in} occupies a central position that can be related to many other conjectures, such as the absence of global symmetries,  the Weak Gravity Conjecture (WGC) \cite{ArkaniHamed:2006dz} and the de Sitter (DS) conjectures\cite{Obied:2018sgi, Ooguri:2018wrx}.  Specifically, the SDC claims that when approaching an infinite distance limit in a field space, an infinite tower of states \footnote{It also could be more than one such towers.} must be generated whose mass $m$ scales exponentially with respect to a Planck scale $M_{Pl}$ in terms of a proper field distance $\Delta D$ as $\frac{m}{M_{Pl}}\sim \text{exp}(-c \Delta D)$, where $c$ is a positive constant and further conjectured as $\mathcal{O}(1)$ in the Refined SDC \cite{Klaewer:2016kiy, Blumenhagen:2018nts}. This conjecture recently has been heavily discussed, elaborated and verified firmly in various contexts of string compactifications \cite{Klaewer:2016kiy, Grimm:2018ohb, Andriolo:2018lvp, Blumenhagen:2018nts, Lee:2018urn,Heidenreich:2018kpg, Lee:2018spm, Gonzalo:2018guu, Lee:2019wij, Blumenhagen:2019qcg, Grimm:2019wtx, Lee:2019xtm, Lee:2019tst, Corvilain:2018lgw, Grimm:2018cpv, Joshi:2019nzi, Baume:2019sry, Marchesano:2019ifh, Grimm:2019ixq, Font:2019cxq, Enriquez-Rojo:2020pqm, Cecotti:2020rjq, Gendler:2020dfp, Andriot:2020lea, Conlon:2020wmc}. The appearance of such an infinite tower of asymptotically massless states, typically indicates that the original effective description must break down at infinite distance limits (where $\Delta D \rightarrow \infty$), and instead, a new description takes over in terms of these asymptotically massless states. Furthermore, the SDC has been refined eloquently as the Emergent String Conjecture in \cite{Lee:2019wij} such that if an infinite distance limit in classical field space falls into the category of the so-called "equi-dimensional" infinite limits\footnote{The "equi-dimensional" infinite distance limit refers that at this limit, the parametrically leading infinite tower of massless states must not be any Kaluza-Klein (KK) or KK-like towers. If instead, the leading tower of massless states comes from the KK-like towers at an infinite distance limit, then it leads to a decompactification (gravity hence is decoupled at the defined dimensional spacetime) and is truly in the disguise of a higher dimensional theory at the limit.  For more details on this concept, we refer to \cite{Lee:2019wij}.}, then the new description reduces to a weakly coupled, fundamental tensionless string theory, where the infinite tower of the asymptotically massless states is furnished by the excitations of the emergent fundamental tensionless string. It was also studied in \cite{Lee:2019wij} that the emergent tensionless fundamental string in the moduli space of 4d $N=2$ vector multiplets receives quantum corrections and the classical "equi-dimensional" limit is obstructed by quantum effects, leading to a decompactification where Seiberg-Witten field theory emerges at the limit instead. Later on, the Emergent String Conjecture has also been discussed and verified nicely in the hypermultiplet moduli spaces of 4d $N=2$ effective theories of type IIB and type I compactifications\cite{Baume:2019sry}.

However, most of the discussions have been done in the context of 4d $N=2$ theories (or their equivalents with eight supercharges) and it might lead to the question of whether the above statements might be resulted from special structures required by theories with eight supercharges. Hence it is natural to explore them in theories with fewer supersymmetries or, even without any supersymmetries, although it is more complicated in general. It is exactly the goal of this paper to explore the SDC, or the refined one- the Emergent String Conjecture, in the context of the 4d $N=1$ effective theories arising from M-theory compactifications. 

In this note, we would like to consider M-theory compactifications on Twisted Connected Sum (TCS)  $G_2$ manifolds. A TCS $G_2$ manifold $X$ can be globally viewed as a $K3$ surface fibered on a three-fold $S^3$ and we consider an infinite distance limit in the moduli space characterized by the shrinking $K3$-fiber on $X$.  We will argue that a weakly coupled, tensionless heterotic string does emerge at this infinite distance limit. Unlike the 5d $N=1$ theories analyzed in \cite{Lee:2019wij}, we would argue that such an emergent tensionless string necessarily receives quantum corrections and hence the tension of the string would not vanish at the infinite distance limit, which also fits with the spirit of the Emergent String Conjecture \cite{Lee:2019wij}. As a by-product, we claim all the tensionless heterotic solitonic strings arising from wrapped D3-branes in F-theory compactifications studied in \cite{Lee:2019tst} regain non-zero tensions at infinite distance limits. Geometrically, we expect that the quantum corrections modify the classical moduli space such that the geodesic trajectory towards to the infinite distance limit is bent, and the infinite distance limit is obstructed at the quantum level, inspired by $N=2$ examples studied in \cite{Marchesano:2019ifh, Baume:2019sry}. 

Our results are also in line with the study \cite{Lee:2019wij} of Type IIA compactification on a K3-fibered Calabi-Yau three-fold $X_3$. In fact, one can view the 4d $N=2$ Type IIA compactification as a weakly coupling limit of M-theory compactification on the manifold $X_3\times S^1$, where $S_1$ denotes the M-theory one-cycle controlling the type IIA string coupling $g_{IIA}$.  Relevant to us, the manifold $X_3\times S^1$ can be treated as a special $G_2$ manifold with the holonomy group being smaller than $G_2$.  And it has been quantitatively studied in \cite{Lee:2019wij} that even this theory with higher supersymmetries, the volume of the various dimensional cycles receives quantum corrections, and hence the classical infinite distance limit, characterized by the shrinking $K3$-fiber on $X_3$, is obstructed in the quantum moduli space of K\"ahler vector multiplets of type IIA.

This note is organized as follows: In section \ref{G2compac} we review the effective action of general $G_2$ compactifications of M-theory and the connections to their weak coupling type IIA orientifold limits, as well as some basic aspects of TCS $G_2$ manifolds. In section \ref{infidistanceweak}, an infinite distance limit is introduced and we argue that the gauge coupling of a certain gauge field becomes weak at such the infinite distance limit. In section \ref{Emergenceoftensionless}, we argue that a weakly coupled, fundamental tensionless heterotic string emerges from such an infinite distance limit, and the excitations of the tensionless string indeed furnish the parametrically leading tower of asymptotically massless states. Quantum corrections on the emergent tensionless string would be discussed in section \ref{quantumcorrections} and we expect the quantum corrections to modify the moduli spaces such that the infinite distance limit is obstructed at the quantum level.  We conclude in section \ref{conclusionoutlook} and also comment on other possible types of infinite distance limits in TCS $G_2$ compactifications, which might lead to an emergence of fundamental type II strings. In appendix \ref{typeIIAorien}, we give a light overview of type IIA Calabi-Yau compactifications and orientifold compactifications, which are relevant for our discussions in the main context.

\section{ Compactifications of M-theory on $G_2$ Manifolds}
\label{G2compac}
\subsection{The Effective Action of $G_2$ Compactifications }
In this section, we give an overview of the effective theories from $G_2$ compactifications of M-theory and set the background for the later discussions in this note.  Many aspects of $G_2$ compactifications have been studied before, for examples in \cite{Kachru:2001je, Cvetic:2001kk, Witten:2001uq, Atiyah:2001qf, Acharya:2004qe, Halverson:2015vta, Halverson:2014tya, Guio:2017zfn, Acharya:2018nbo, Andriolo:2018yrz}.

A $G_2$ manifold is a real seven-dimensional Riemannian manifold $ X$ with a Riemannian metric $g$ such that the holonomy group $\text{Hol}(X)\subseteq G_2$. Physically speaking, with such a holonomy group $\text{Hol}(X)\subseteq G_2$, it leads to a global covariantly constant spinor $\eta$ as the spinor representation $\bf 8$ of $SO(7)$ decomposes as 
\be
\bf 8 \rightarrow \bf7+\bf1.
\ee 
One can use the spinor $\eta$ to construct covariantly constant three-form $\Phi$ and its hodge dual four-form $\ast_X \Phi$, dubbed $associative$ form and $coassociative$ form, which are invariant under the holonomy group $\text{Hol}(X)$. The $associative$ form $\Phi$ and $coassociative$ form $\ast_X \Phi$ are further calibration forms, which can calibrate minimal volumes of the corresponding three-cycles and four-cycles, respectively.  The three-form $\Phi$ can also be used to determine a Riemannian metric $g$ on $X$. The first examples of compact $G_2$ manifolds were constructed by D. Joyce \cite{Joyce:1996dr} from resolutions of the orbifolds $T^7/\Gamma$ (or $(K3\times T^3)/\Gamma$) with $\Gamma$ being suitable finite groups. 

The compactifications of M-theory on $G_2$ manifolds typically lead to  theories of 4d $N=1$ supergravity, which can be determined by three functions: a F-term superpotential $W$, a K\"ahler  potential $K$ and a gauge-kinetic coupling function $f_{\alpha\beta}$. We are going to briefly summarize these three functions in terms of geometry of $G_2$ manifolds  following \cite{Grimm:2004ua}.

To start with, we recall that the relevant parts in the 11d supergravity effective action is given by 
\be
S_{11d}=\frac{1}{\kappa^2_{11}} \int_{R^{1,10}}(\sqrt{-g}R-\frac12d C_3\wedge \ast dC_3-\frac16 C_3\wedge dC_3\wedge dC_3).
\ee
with $\kappa^2_{11}:=\frac{\ell_{11}^9}{2\pi}$.
After compactified on a $G_2$ manifold $X$, the first term tells us the 4d Planck mass $M_{Pl}$ is given by the volume of $X$ as
\be
\frac{M_{Pl}^2}{M_{11}^2}=4\pi \text{Vol}(X)
\ee

The Kaluza-Klein ansatz for the M-theory $C_3$ field, complexfied by the M-theory three-form $\Phi$, reads
\be
i\Phi+C_3=i\sum_{i=1}^{b_3}\phi_i s^i+A^\alpha \omega_\alpha ,, \qquad \phi_i \in H^3(X), \omega_\alpha \in H^2(X),
\ee
from which one can infer that the effective theory has $b^3(X)$ chiral multiplets built from the decomposed fields $s^i$, together with $b^2(X)$ vector multiplets built from the fields $A_\alpha$.

The metric of moduli space is determined by the K\"ahler potential, which at the large volume regime in the moduli space of $X$ can be expressed as  
\be
K^{G2}=-3\rm{log} (\frac{1}{7\kappa_{11}^{2}}\int_X\Phi\wedge \ast_X \Phi),
\ee
where $V_c(X):=\frac1{7}\int_X\Phi\wedge \ast_X \Phi$ exactly represents the (classical) volume of the $G_2$ manifold $X$. The associated K\"ahler metric hence is given by 
\be
G_{i\bar j}=\partial_i\bar{\partial}_{\bar j}K^{G_2}=\frac14V_c(X)^{-1}\int_X\phi_i\wedge \ast_X \phi_j.
\ee
As a well-known fact, there is a no-scale structure associated with the 4d potential $V$, which results from
\be
\partial_i K^{G_2}G^{i\bar j} \bar{\partial}_{\bar j}K^{G_2}=7,
\ee
where $G^{i\bar j} $ is the inverse of the K\"ahler  metric $G_{i\bar j}$. Notice that the K\"ahler potential receives further quantum corrections, mainly from M2-brane instantons, which breaks the no-scale structure when such corrections are not highly suppressed.

The gauge-kinetic coupling function $f_{\alpha\beta}$ is given by
\be
f_{\alpha\beta}=\frac{i}{2\kappa_{11}^2} \int_{X}\omega_{\alpha}\wedge\ast_X \omega_{\beta}.
\ee

Finally, non-vanishing background $G_4=dC_3$ fluxes induce a superpotential term
\be
W_{G_2}=\frac{1}{4\kappa_{11}^2}\int_X(\frac12C_3+i\Phi)\wedge G_4.
\ee
Note that M2-instantons from wrapping M2-branes on rigid $associative$ cycles also induce a non-perturbative superpotential. However, we would not consider all F-terms in the later discussions, but leave that for a further investigation in the future.

Before closing this subsection, we would like to briefly comment on connections to type IIA orientifolds, if such $G_2$ compactification exists Type IIA orientifold limits, see more details of Type IIA orientifold compactifications in \ref{typeiiaorientifoldcompactific}. It is well-known that Type IIA orientifold with $O6^-$-planes without any RR fluxes can be lifted to $G_2$ orbifolds \cite{Kachru:2001je} as
\be
\label{typeIIAupliftg2}
X=(X_3\times S^1)/(\sigma, -1),
\ee
where $X_3$ is a Calabi-Yau manifold in Type IIA compactifications and $S^1$ is the M-circle whose radius $R$ controlling the type IIA string coupling. In general, we have the following relations between the energy scales measured in the two different frames \cite{Witten:1995ex}, 
\be
\label{scalestypeiiaM}
g_{IIA}= (M_{11} R)^{3/2}, \qquad   R=g_{ IIA}\ell_{s},
\ee
where $M_s:=\frac1{\ell_{s}}$ denotes the 10d type IIA string scale and $g_{ IIA}$ is the type IIA string coupling. For simplicities, we set the 11d Planck scale to be $M_{11}:=\ell_{11}^{-1}=1$ in the rest of this note unless it is needed.

In this scenario, the $G_2$ three-form $\Phi$ can be expressed in terms of K\"ahler two-form $J $ and Calabi-Yau three-form $\Omega^{3,0}$ on $X_3$, which systematically reads \footnote{$J_M, \Omega_M^{3,0}$ here denotes a rescaling version of $J, \Omega^{3,0}$ in the string frame, see more details on the notations in \cite{Grimm:2004ua}.}
\be
\label{threeform}
\Phi= J_M\wedge dx+\text{Re}(\Omega_M^{3,0}), \qquad \ast_X \Phi=\frac12 J_M\wedge J_M+\text{Im}(\Omega_M^{3,0})\wedge dx.
\ee
where $dx$ is the non-trivial one-form on $S^1$.
And one can verify that 
\be
K^{G2}=-\text{ln}(-\frac43\int_{X_3}(J\wedge J\wedge J))-2\text{ln}(2\int_{X_3}\text{Re}(C\Omega)\wedge \ast \text{Re}(C\Omega)):=K^{IIA}.
\ee
where $K^{IIA}$ denotes the K\"ahler potential in type IIA orientifold compactifications, introduced in \ref{typeiiaorientifoldcompactific}.

Finally, a $G_2$ manifold $X$ can have non-trivial cycles $H_i(X, \mathbb R), i=2,3,4,5$. Relevant to us, we would like to mention the correspondences of the cohomologies in a orientifold $X_3/\sigma$ and the ones in a $G_2$ manifold $X$ in general, which are given by 
\be
\label{hodgedecom}
\begin{split}
H^2(X)=H^2_+(X_3), \qquad &H^3(X)=H^3_+(X_3)\oplus[H^2_-(X_3)\wedge H^1_-(S^1)] \\
H^4(X)=H^4_+(X_3)\oplus[H^3_-(X_3)\wedge H^1_-(S^1)],\qquad &H^5(X)=H^4_-(X_3)\wedge H^1_-(S^1),
\end{split}
\ee
where $H^1_-(S^1)$ denotes the odd one-form of the M-circle $S^1$ under the involution \eqref{typeIIAupliftg2}. Such cycles, depending on the dimensions, support various physical objects such as particles, strings and domain-walls from wrapped M2-branes and M5-branes.

For the rest of this note, we reserve the notation $X_3$ for a generic (K3 fibred) Calabi-Yau three-fold and $X$ for a generic (TCS) $G_2$ manifold. 

\subsection{TCS $G_2$ manifolds, Duality and Type IIA Orientifold Limits}
From the above subsection, one can see that the physical context of the 4d effective theories is dictated by the moduli spaces of $G_2$ manifolds, which is analogous to Calabi-Yau compactifications in many ways.  However, comparing to Calabi-Yau manifolds where the moduli spaces have been extensively studied, the study of the moduli spaces of $G_2$ manifolds turn out to be quite difficult and a description at the level of details available for Calabi-Yau cases is still missing. Partly, this is because of the lack of an analog of "Yau's" theorem to guarantee the existence of a $G_2$ metric. Furthermore, $G_2$ manifolds are real and hence the powerful machinery of complex algebraic geometry cannot be applied. In this note, we mainly study one special type of $G_2$ manifold constructed through a twisted connected sum by a combined use of string dualities and weakly coupled type IIA orientifold limits. 

In order to make the context self-contained, we would like to briefly summarize the constructions of TCS $G_2$ manifolds and their weak coupling limits of type IIA orientifolds. For details, we refer to the original mathematics literatures \cite{kovalev2000twisted, Corti:2012kd, Corti2013} and the physics references \cite{Braun:2016igl,Braun:2017ryx, Braun:2017uku, Braun:2017csz, Braun:2018vhk, Braun:2019lnn, Braun:2019wnj}. We will adapt the same notations for the constructions of TCS $G_2$ manifolds with \cite{Braun:2019wnj}.

 TCS $G_2$ manifolds are the special compact $G_2$ manifolds that can be constructed by a twisted connected sum of two "building blocks" $Z_\pm$, together with a product of one-circle $S^1_{e\pm}$ for each of them, respectively. Each of the two building blocks $Z_\pm$ can be viewed as a K3-fibration over an open $\mathbb P^1$ as
 \be
 Z_\pm :=K3 \rightarrow \mathbb P^1,
 \ee
 with the first Chern class being $c_1( Z_\pm)=[S_\pm]$, where $[S_\pm]$ denotes the class of a generic $K3$-fiber among a lattice polarized family of $K3$ surfaces. Excising a generic fiber  $S^0_\pm$ from $Z_\pm$, the remaining parts, denoted as  
 \be
 X_\pm=Z_\pm \backslash S_\pm^0,
 \ee 
 can be viewed as a pair of asymptotically cylindrical (acyl) Calabi-Yau three-folds, each of which is diffeomorphic to the product of a K3 surface $S^0_{\pm}$ and a cylinder $S_{b\pm}^1\times I$ outsider a compact sub-manifold, respectively. 
 
 Then, a TCS $G_2$ manifold $X$ can be constructed by gluing $X_\pm\times S^1_{e\pm}$ along the excised region in a way that $S^1_{e\pm}$ are identified with $S^1_{b\mp}$, respectively and together with a hyperk\"ahler (Donaldson) rotation $\phi$ on the two $K3$-fibers by mapping $S_+^0$ to $S_-^0$ as
 \be
 \phi: \omega_{S^0_{\pm}}\leftrightarrow \text{Re}(\Omega^{2,0}_{S^0_{\mp}}), \qquad \text{Im}(\Omega^{2,0}_{S^0_{\pm}})\leftrightarrow -\text{Im}(\Omega^{2,0}_{S^0_{\mp}}),
 \ee
 where $\omega$ denotes the K\"ahler form on the two $K3$-fibers $S^0_{\pm}$ and $\Omega^{2,0}$ refers to the holomorphic two-form on $S^0_{\pm}$, describing their complex structure moduli. 
 
 Globally, a TCS $G_2$ manifold $X$ can be viewed as a $K3$ fibration over a three-fold $S^3$ \footnote{In this note, we consider the $K3$-fiber on $X$ also carries an elliptical fibration.}, which can be pictured from the fiber-wise duality between M-theory on a $K3$ surface and ($E_8\times E_8$) Heterotic string on a $T^3$ manifold:
\be
\ba
\label{projectionmap}
K3\longrightarrow & X   \qquad \qquad T^3\longrightarrow  X_3  \cr
& \bigg\downarrow  \qquad \qquad\qquad\qquad\bigg\downarrow \cr
&S^3   \qquad \qquad \qquad \quad S^3.
\ea
\ee
  
  The $K3$-fiber is conjectured to be a co-associated four-cycle on $X$, which plays a similar role of the special Lagrangian $T^3$-fibre in a Calabi-Yau three-fold $X_3$ at the large complex structure limit.  Indeed, the above duality has been checked in detail in \cite{Braun:2017uku} by identifying the same massless spectra on both sides. On the dual heterotic side, the SYZ fibration Calabi-Yau three-fold $X_3$ turns out to be the Schoen's Calabi-Yau three-fold $X_{19,19}$, known as the split bi-cubic, where the subscript denotes its Hodge numbers being $h^{1,1}(X_{19,19})=h^{2,1}(X_{19,19})=19$.  The choices of the bundles on $X_{19,19}$ correspond to the choices of the singularities of the $K3$-fiber on $X$\cite{Braun:2017uku}. 
   
 Note that the Schoen's Calabi-Yau three-fold $X_{19,19}$ has another description, which can be viewed as an elliptic fibration over a "del Pezzo-nine" surface $dP_9$. By Heterotic/F-theory duality, one can then extend to F-theory via replacing the elliptic curve on $X_{19,19}$ by an elliptic K3 surface, with fiber $\mathbb E$. The resulted Calabi-Yau four-fold turns out to be the Grass-Donagi-Witten manifold $X_4$: elliptic fibration over the base $B_3:=\mathbb P^1 \times dP_9$ \cite{Donagi:1996yf}. Note that any M-theory compactification on a TCS $G_2$ manifold of the type considered in this note, i.e. with the $K3$-fiber being elliptic, is dual to Heterotic string on $X_{19,19}$ and hence also dual to F-theory on $X_4$ \cite{Braun:2017uku}. We refer to \cite{Braun:2017uku, Braun:2018fdp} for more details on these dualities.  All-in-all we have the following duality chain:  
 \be
 \label{stringduality}
 \text{M-theory on a $G_2$ X}  \Longleftrightarrow \text{Heterotic string on}~ X_{19,19} \Longleftrightarrow  \text{F-theory on}~ X_4.
 \ee

   For the later purpose, we particularly point out the following relation under the above duality chain:
\be
\label{hertodilat}
 \frac{e^{3\gamma}}{\text{Vol}(S^3)}=g_{het, 4d}^2=\frac{\text{Vol}(\mathbb P^1)}{\text{Vol}(dP_9)},
\ee
where $e^\gamma$ denotes the radius of the $K3$-fiber with the total volume being normalized as $\text{Vol}(K3)=e^{4\gamma}$. To see that, one recall that the 8d heterotic string coupling is given by $ g_{het, 8d}^2=\text{Vol}(\mathbb P^1)$ in the duality between F-theory on elliptically fibered $K3$ (whose base is $\mathbb P^1$) and Heterotic string on $T^2$ \cite{Vafa:1996xn}. Compactified them further on $dP_9$ to the 4d effective theories, the 4d heterotic string coupling $g_{het, 4d}$ hence follows that 
\be
g_{het, 4d}^2=\frac{\text{Vol}(\mathbb P^1)}{\text{Vol}(dP_9)}.
\ee
Regarding the left equality, one can first lift to 7d and recall that the heterotic string coupling $g_{het, 7d}$ is given by $e^{3\gamma}$ \cite{Witten:1995ex} \footnote{Alternatively, one can also verify by using standard M/F-theory duality that it is $e^{3\gamma}$ rather than the volume of the $K3$-fiber $e^{4\gamma}$. Namely, by dualising into F-theory on the $K3$-fiber, the heterotic string coupling $g^2_{het, 8d}$ is determined by the size of the base $\mathbb P_b^1$ of the $K3$-fiber, i.e. $g^2_{het, 8d}=\text{Vol}(\mathbb P_b^1)$. Compactified it further on $S^1$ with the radius being $r$, it gives rise to $g^2_{het, 7d}=\frac{g^2_{het, 8d}}{r}=\frac{\text{Vol}(\mathbb P^1)}{r}$. Noting that the radius $r$ under standard F/M-duality amounts to $\text{Vol}(\mathbb E)^{-1/2}$, we then have 
 \be
 g^2_{het, 7d}=\text{Vol}(\mathbb E)^{1/2} \text{Vol}(\mathbb P^1) \neq \text{Vol}(K3).
 \ee
} in the duality between M-theory on $K3$ and Heterotic string on $T^3$. Further compactified them on the 3-cycle $S^3$ on $X$ to 4d, one has $ g^2_{het, 4d}=\frac{ g^2_{het, 7d}}{\text{Vol}(S^3)}$ which produces the result.

Now we switch gear to briefly discuss the type IIA orientifold limits of a TCS $G_2$ manifold. A recent nice paper which has been done regarding this aspect appears in \cite{Braun:2019wnj}, and we refer to it for more details on the connections. Here we briefly summarize the relevant aspects. It has been shown in \cite{Braun:2019wnj} that the existence of a type IIA orientifold limit for such a TCS $G_2$ manifold $X$ requires that the Calabi-Yau three-fold $X_3$ in \eqref{typeIIAupliftg2} prior to the orientifolding should carry a $K3$ fibration and the anti-holomorphic involution $\sigma$ should be respected the $K3$ fibration on $X_3$ and hence the fibration can be lifted to $X$. More precisely, the two building blocks of the TCS $G_2$ $X$ after a resolution of the orbifold \eqref{typeIIAupliftg2} take as following: the building block $Z_-$ inherits the $K3$-fiber of $X_3$ and fibers it over an $\mathbb P^1$, which further is a half Calabi-Yau three-fold $X_3$, whereas the other building block $Z_+$ is a "Voisin-Borcea" type obtained from the Voisin-Borcea Calabi-Yau three-fold. %The M-theory circle $S^1$ is given by the $S_{e^-}^1$, which becomes a part of the base of K3 fibration $X_{+}:=Z_+\backslash S_+^0$ and is not of constant size, as expected. 

  From the perspective of phenomenology, it appears to be trivial to consider any M-theory compactifications on TCS $G_2$ manifolds as they seems do not have codimension-seven singularities, hence the chiralities of the 4d effective theories are trivial \footnote{ A novel proposal for generating non-trivial chiralities, however, has been made recently in \cite{Barbosa:2019bgh} by exploiting the T-branes-like in TCS $G_2$ manifolds. }. Nevertheless, such geometries provide many interesting insights on many other aspects such as \cite{Braun:2018vhk, Halverson:2014tya, Halverson:2015vta}.

\section{Infinite Distance Limits as Weak Gauge Coupling Limits}
\label{infidistanceweak}
As mentioned in the previous section, a TCS $G_2$ manifold $X$ can be globally viewed as a $K3$ fibration over a three-fold $S^3$. The infinite distance limit in the moduli space of M-theory on $X$ we are interested in, is characterized by the infinite volume of the base $S^3$ and the zero size of the $K3$-fiber, together with the condition that the total volume of TCS $G_2$ manifolds stay fixed, in order to keep the 4d Planck scale $M_{Pl}^2$ finite and hence gravity is not decoupled in 4d \footnote{Note that such a limit has also been studied in the mathematics literature \cite{donaldson2016adiabatic}, dubbed an adiabatic limit. }. Quantitatively, we can assign a scaling $\mu$ as 
\be
\label{infinitlimit}
\begin{split}
\text{Vol}(K3) &\rightarrow \mu^{-1} \text{Vol}(K3), \quad  \text{Vol}(S^3) \rightarrow \mu \text{Vol}(K3),\qquad  \text{with} ~ \mu \rightarrow \infty  \\
&\text{Vol}(X) \propto \text{Vol}(K3)\times \text{Vol}(S^3)\sim \text{finite} .
\end{split}
\ee
As alluded above, it is hard to prove that such limit is at the infinite distance in the moduli space of $X$ given that we know little about the moduli space and the metric $g_{\alpha\beta}$, in contrast to the detailed study in Calabi-Yau compactifications. However, recall that such a TCS $G_2$ compactification is dual to F-theory on $X_4$ and from \eqref{hertodilat}, we learn the above infinite limit can be mimicked in the F-theory frame as 
\be
\label{infinitlimit2}
\begin{split}
\text{Vol}(\mathbb P^1) &\rightarrow \nu^{-1} \text{Vol}(\mathbb P^1), \quad \text{Vol}(dP_9) \rightarrow \nu \text{Vol}(dP_9),  \qquad \text{with} ~\nu \rightarrow \infty \\
&\text{Vol}(B_3) \propto \text{Vol}(\mathbb P^1)\times \text{Vol}(dP_9)\sim \text{finite},
\end{split}
\ee
where $\nu$ denotes the corresponding scaling in the dual F-theory frame. And such a limit is indeed at infinite distance in the moduli space of F-theory on $X_4$, as an example of the generic cases of the bases $B_3$ being $\mathbb P^1$ fibration over Fano two-folds proved in \cite{Lee:2019tst}. By merit of string duality, we hence claim that the limit \eqref{infinitlimit} is indeed at the infinite distance in the moduli space of $X$.  

Furthermore, if a $G_2$ manifold $X$ has a type IIA orientifold weak coupling limit, and since the $K3$-fibre should descend to the type IIA orientifold $X_3/\sigma$, then the corresponding infinite distance limit is naturally mimicked by the shrinking $K3$-fiber in the orientifold.The Swampland Distance Conjecture in type IIA orientifold compactifications has recently been discussed in \cite{Font:2019cxq}, where they generalized the tower of asymptotically massless particles to towers of asymptotically massless branes (as higher-dimensional objects), without considering any quantum corrections.

The infinite distance limits in the field space of an effective theory are typically expected to also be weak gauge coupling limits associated with relevant gauge fields, which in our cases arise from decompositions of $C_3$ along two-cycles on $G_2$ manifolds. To see that, we recall a gauge coupling $g_{YM} $ is given by 
\be
\frac1{4g_{YM}^2}=\text{Re}f_{\alpha\beta}=\text{Re}(\frac{i}{2\kappa_{11}^2} \int_{X}\omega_{\alpha}\wedge\ast_X \omega_{\beta}).
\ee

We would like to show that the gauge kinetic function $g_{\alpha\beta}$ tends to zero in the points where certain three-cycles tend to infinite sizes. To this end, we recall that for a $G_2$ manifold $X$, whose holonomy is exactly $G_2$ group, then its second cohomology is all in the $\mathbf{14 }$ representation of a $G_2$ group, leading to $H^2(X,\mathbb R)=0$. As a result, we can write the hodge dual as 
\be
\ast_{X} w_{\alpha}=-w_{\alpha}\wedge \Phi.
\ee
Hence, we can rewrite the above gauge coupling as a volume of a three-cycle $D_\Sigma$ \cite{Halverson:2015vta} as 
\be
\begin{split}
\frac1{4g_{YM}^2}:=\text{Re} f_{\alpha\beta} =&\text{Re}(\frac{i}{2\kappa_{11}^2} \int_{X}\omega_{\alpha}\wedge\ast_X \omega_{\beta})\\
\propto&\int_{X} w_{\alpha}\wedge w_{\beta}\wedge\Phi =Vol(D_\Sigma),
\end{split}
\ee
where $D_\Sigma$ is a (putative) three-cycle whose class is dual to the class of the four-form $\omega_\alpha\wedge \omega_{\beta}$.  Such a canonical map can be viewed as a result from Poincar\'e duality with homology and the Joyce lemma \cite{Halverson:2015vta}.  

 From the above, we hence draw an conclusion that a weak gauge coupling limit can be achieved if there is a three-cycle whose volume tends to infinite, 
\be
\text{Vol}(D_\Sigma) \rightarrow \infty \Longrightarrow g_{YM}^2 \rightarrow 0,
\ee
 where our infinite distance limit \eqref{infinitlimit} directly applies when taking the three-cycle $D_\Sigma$ as the base $S^3$ on $X$. This is similar in spirit to cases in the Calabi-Yau compactifications \cite{Grimm:2018ohb, Corvilain:2018lgw, Lee:2018urn, Lee:2018spm, Lee:2019wij}. The above is also consistent with the well-known conjecture in ($d\geqslant 3$) quantum gravity: there are no global symmetries. Indeed, as alluded above, when a gauge coupling $g_{YM}$ tends to zero, it implies that the corresponding gauge symmetry becomes a global symmetry, which is prohibited when coupled to gravity according to the conjecture, and hence such a limit should not be attained within the original effective theory. Therefore, such a limit must lie at the infinite distance in a field space where the original effective description breaks down, due to the appearance of infinite towers of asymptotically massless states, which is a necessary result of the Swampland Distance Conjecture.  In \cite{Grimm:2018ohb, Corvilain:2018lgw}, they further clarified that any infinite tower of massless states can be viewed as quantum gravity obstructions to restore a global symmetry at the infinite distance limits.

\section{Emergence of the tensionless fundamental heterotic string}
\label{Emergenceoftensionless}
After introducing the infinite distance limit \eqref{infinitlimit} in the moduli space of M-theory compactification on $X$, we are going to discuss the possible candidates for the leading infinite tower of asymptotically massless states predicated by the SDC.  

As we expected from String/M-theory compactifications, there are typically two sources for these infinite towers: KK towers and stringy towers \cite{Palti:2019pca}. KK towers are universally present in any compactifications. Regarding the asymptotically massless stringy towers in our cases, we consider an M5-brane wrapping the shrinking $K3$-fiber at the limit \eqref{infinitlimit} and leads to a tensionless string in the 4d effective theories. Such a wrapped M5-brane typically preserves $N=(0,2)$ supersymmetry on the 2d world-volume \cite{Gadde:2013sca, Benini:2013cda} and turns out to be a fundamental heterotic soliton string in 4d \footnote{Namely, this wrapped M5-branes (as a soliton solution of the bulk supergravity) can be identified to a fundamental heterotic string probing the dual heterotic geometry and can be described by a heterotic world-sheet string theory.}. This can be viewed as a fiber-wise application of the results in \cite{Harvey:1995rn, Cherkis:1997bx}, and should not come as a total surprise but manifest the Heterotic/M-theory duality. Indeed, one can alternatively perform a dimensional reduction along a generic $K3$ surface from the 6d $N=(0,2)$ world-volume theory of an M5-brane, and find that the 2d effective theory on the string exactly produces 8 right-moving massless scalars together with its supersymmetric partner fermions, and 16 left-moving massless fermions and 8 left-moving massless scalars, which fits with the massless spectra of a weakly coupled heterotic fundamental string. Embedding the K3 surface into $X$, the supersymmetry on the 2d effective theory reduces to $N=(0,2)$. The same spectra and supersymmetry can also be verified on the 2d string in the dual F-theory frame under the duality \eqref{stringduality}, which arises from a D3-brane wrappping on the curve $C_0:=\mathbb P^1$ on $X_4$ \cite{Lawrie:2016axq, Lee:2019tst} (See the table $\bf 5$ in \cite{Lawrie:2016axq} with $C_0 \cdot \bar {K}_{B_3}=2$ and trivial normal bundle $N_{C_0/B_3}=\mathcal O_{C_0}\oplus \mathcal O_{C_0}$ in our case).

One can also see that this tensionless fundamental heterotic string is weakly coupled \footnote{This tensionless fundamental string should be contrasted with a non-critical tensionless string, which typically arises from wrapping a M5-brane on a contractible four-cycle at some finite distances in the moduli space and instead indicates the appearance of a strongly coupled superconformal theory. } at the limit \eqref{infinitlimit}, as the heterotic string coupling $g_{het}$ is determined in \eqref{hertodilat}. 
The excitations of this weakly coupled, tensionless fundamental heterotic strings lead to a tower of asymptotically massless states, whose mass at the $n, n=0,1,...,\infty$ level scales as
\be
\label{stringytower}
\frac{M_{n, Het}^2}{M_{Pl}^2} \propto n \frac{T^2_{Het}}{M_{Pl}^2}= n\frac{2\pi\text{Vol}(K3) M_{11}^2}{M_{Pl}^2}\propto n\frac{M_{11}^2}{\text{Vol}(S^3)}\propto \frac{n}{\mu}, ~ ~ \mu \rightarrow \infty,
\ee 
where $\frac{T_{Het}}{M_{11}^2}:=2\pi\text{Vol}(K3) $ denotes the tension of the emergent heterotic string in unites of 11d Planck mass $M_{11}$, and $\mu$ denotes the scaling in \eqref{infinitlimit}. In the third step, we have used $\text{Vol}(K3)\propto \frac{\text{Vol}(X)}{\text{Vol}(S^3)}$ and $\text{Vol}(S^3) \propto \mu$ at this infinite distance limit and $M_{Pl}^2=2\pi\text{Vol}(X) $, in the same fashion as the $K3$ fibration Calabi-Yau three-fold discussed in \cite{Lee:2019wij}. 

While the competing Kaluza-Klein tower at the $n$ level scales as 
\be
\label{KKtower}
\frac{M_{n,kk}^2}{M_{Pl}^2} \propto \frac{n^2}{\text{Vol}(S^3)M_{Pl}^2}\propto \frac{n^2}{\mu} \frac{1}{\text{Vol}(X)}, \qquad \mu \rightarrow \infty,
\ee
where we have used the fact that the scale of the KK tower is determined by the inverse volume of the largest cycle in a compactified space, which is the base $S^3$ on $X$.

As one can see now, the KK tower is at the same scaling level with the stringy tower \eqref{stringytower} generated by the tensionless fundamental heterotic string, but it is less dense as it is proportional to $n^2$ rather than $n$, hence we claim that the limit \eqref{infinitlimit} is indeed an "equi-dimensional" limit \cite{Lee:2019wij} and does not lead to a decompactification, at least classically.  This fits with the spirit of the Emergent String Conjecture. Physically, this means that the original effective theory breaks down at the infinite limit due to the infinite tower of massless states, and reduces to an effective theory of weakly coupled, asymptotically tensionless heterotic string as it provides the leading tower.

Before we continue to next topics, we would like to point out one point regarding the other possible candidate of an infinite tower of asymptotically massless states, which arises from wrapping an M2-brane on a non-contractible curve $C_0$ inside the shrinking $K3$-fiber $m,  m=0,1..., \infty$ times. Recall that in the 5d $N=1$ case from M-theory compactification on a K3-fibered Calabi-Yau three-fold $X_3$, which has been extensively discussed in \cite{Lee:2019tst}, the infinite tower of asymptotically massless BPS states from an M2-brane wrapping $m$ times on a distinguished two-cycle $C_0$ with $C_0\cdot C_0 >0$ on the shrinking $K3$-fiber, corresponds to the dual heterotic string wrapping a one-cycle $S_A^1$ $m$ times on the heterotic dual geometry $\hat S \times S_A^1\times S_M^1$, where $\hat S$ denotes the dual K3 surface to the $K3$-fiber. This can also be verified by exploiting the 4d $N=2$ string duality between type IIA on $X_3$ and heterotic string on $\hat S \times T^2$ \cite{Harvey:1995fq, Kachru:1995wm}. Such BPS states from the wrapped M2-brane are known as Gopakumar-Vafa invariants \cite{Gopakumar:1998ii, Gopakumar:1998jq, Dedushenko:2014nya} of $X_3$. However, with generic smooth TCS $G_2$ manifold, the heterotic dual geometry is the Schoen's Calabi-Yau three-fold $X_{19,19}$, which does not possess a non-trivial holomorphic one-cycle $S_A^1$ as $H^{1,0}(X_{19,19})=H^{0,1}(X_{19,19})=0$. So what happens? At first sight, one may turn to the SYZ fibration for help illustrated in the fiber-wise duality \eqref{projectionmap}. At the large complex structure regime, a Calabi-Yau three-fold $X_3$ can be viewed as a special Lagrangian $T^3$ fibered over a three-cycle. We can see that although the fiber $T^3$ itself has non-trivial one-cycles, the class of such cycle becomes trivial measured by the complex structure of $X_3$.  Inspired by this, one would naturally expect that the two-cycle $C_0$ should not exist on the TCS $G_2$ manifold, namely although the $C_0$ is a non-trivial two-cycle on the $K3$-fiber of $X$, it would not be so on the whole $G_2$ manifold.

However, the above argument fails when the $K3$-fiber of $X$ is algebraic. Instead, a $K3$-fiber being algebraic in the $X$ means that there is at least one non-trivial two-cycle on $X$ that is inherited from the $K3$-fiber. To see that, recall that the Picard lattice of an algebraic $K3$ surface reads
 $$\text{Pic}(K3)=H^{1,1}\cap H^2(K3, \mathbb Z), $$
 which is a lattice with the signature $(1, \rho-1), 1\leqslant \rho \leqslant 20$, see for example \cite{Aspinwall:1996mn} for more details.  One can define a map 
\be
a: K3 \rightarrow  X,
\ee
which can induce a lattice from the pull-forward $a^\ast$ as
\be
\Lambda=[a^\ast H_2(X, \mathbb Z)].
\ee
Such lattice  $\Lambda$ must be of signature $(1, r-1), r\leqslant \rho$, which claims at least one non-trivial two-cycle on $X$ that can be obtained by the push-forward $a^\ast$ from the algebraic $K3$ surface. In such a scenario, we expect that the underlying duality makes sense only down to the 2d $N=(2,2)$ theory, given by 
$$
\text{Type IIA on}~ X_4  \leftrightarrow \text{Heterotic string on}~ X_3\times T^2,
$$
where the particles from wrapping an M2-brane on a distinguished curve $C_0$ insider the $K3$-fiber on $X$ corresponds to the ones from wrapping heterotic string on a one-cycle $S^1$ on $T^2$. We will leave it for a future study in a meticulous manner.

We have claimed that a weakly coupled, tensionless fundamental heterotic string indeed emerges at the infinite distance limit \eqref{infinitlimit}, of which the stringy excitations furnish an infinite tower of the asymptotically light states at this limit, which is in line with the Emergent String Conjecture \cite{Lee:2019wij}. However, we have only discussed it so far at the classical level and have not considered any quantum corrections. Indeed, within the 4d $N=1$ effective theories, it is expected that any strings with $N=(0,2)$ supersymmetry, whether being critical or non-critical, generally do not receive protections from supersymmetry against quantum corrections, contrasted with their counterparts in 5d or 6d theories. In general, the study of quantum corrections in 4d $N=1$ effective theories is a difficult problem, even in an attempt of determining the exact form of F-terms. In our cases of $G_2$ compactifications, such a difficulty would be complicated by limited knowledge on the moduli space and how to embed the $associative$ three-cycles into $X$, see for examples \cite{Acharya:2018nbo, Braun:2018fdp} for recent progress on this aspect.  Nevertheless, we do not need full-fledged knowledge of quantum corrections but only need to verify whether or not the quantum corrections, mainly from M2-instantons, obstruct one from approaching the infinite distance limit \eqref{infinitlimit}.

\section{Quantum Corrections}
\label{quantumcorrections}

\subsection{Quantum Volumes}
In this section, we switch gear to study quantum corrections in TCS $G_2$ compactifications of M-theory. Our main goal is to try to verify that the vanishing volume of the $K3$-fiber at the limit \eqref{infinitlimit} receives non-zero contributions from quantum corrections, leading to an obstruction of approaching the infinite distance limit at the quantum level. 

To begin with, we would like to recall that how a similar problem in the 4d $N=2$ settings of Type IIA Calabi-Yau compactifications has been solved in \cite{Lee:2019wij}, where they showed that the volume of the $K3$-fiber in a Calabi-Yau $X_3$ indeed receives quantum corrections such that it never vanishes in the quantum moduli space of the 4d $N=2$ vector multiplets. One useful observation for us is that, one can view the above Type IIA compactification as a special example of $G_2$ compactifications, where the $G_2$ manifold is $X_3\times S_M^1$, of which the holonomy group is $\text{Hol}(X_3\times S^1)=SU(3)\subseteq G_2$. For the sake of our later discussions, we start by briefly summarizing their relevant arguments.  

 The upshot is that the quantum volumes of various dimensional cycles in the Type IIA quantum moduli spaces are truly measured by the underlying world-sheet sigma models and they are expected to receive corrections such as those from world-sheet instantons. Such quantum volumes of various $2n$-cycles $C$ on $X_3$ are only identical with the classical ones measured by the K\"ahler form $J$, i.e. $\text{Vol}_c(C):=\int_{C} J^n, n=1,2,3$, at the large volume regime of the K\"ahler moduli space $\mathcal M^K$, see the notations of 4d N=2 moduli spaces in the appendix \ref{typeIIAorien}. In general, one can use mirror symmetry to determine such quantum volumes by identifying them with the corresponding entries in the period vector \eqref{periodvector}, which are constructed in the mirror Calabi-Yau three-fold dual to $X_3$. Physically speaking, such quantum volumes are also accounted for the physical masses of wrapped B-branes. For example, the quantum volume of the $K3$-fiber on $X_3$ identifies with the mass of the particle arising from wrapping a D4-brane \footnote{To be more precise, the particle should arising from a bound state of one wrapped $D4$-brane and one anti-D0-brane, due to the curvature effect of the $K3$-fiber, see more details in \cite{Lee:2019wij}. } along the $K3$-fiber.  Furthermore, the particle is also identified with an M5-brane wrapping $K3\times S_M^1$ followed the reduction of M-theory to Type IIA, where $S_M^1$ denotes the M-circle with the radius $R$. Now the crucial point is that the wrapped M5-brane on the shrinking $K3$-fiber is exactly the weakly coupled heterotic fundamental string, and hence one can read the mass of the particle by utilizing the mass formula for the fundamental string \cite{Lee:2019wij}.

That is, the total mass of the wrapped heterotic string on $S_M^1$ with the winding number $n=1$ and no KK momentum, identical to the mass of the wrapped D4-brane, is given by
 \be
 \label{massformular}
 \frac{M_n}{M_{11}}=| \frac{ n RT_{H}}{M_{11}}+\frac{E_0}{R{M_{11}}}|, %\qquad n=0,1,..., \infty,
 \ee
 where $E_0$ denotes the Casimir zero energy, which has been shown in \cite{Bobev:2015kza} relates to the gravitational anomaly on the world-sheet theory of the fundamental heterotic string by 
 \be
 E_0=-\frac{\chi(K3)}{24}=-1.
 \ee
Here $\chi(K3)=24$ denotes the Euler characteristic of the $K3$-fiber. Such the non-vanishing Casimir zero energy turns out to be closely tied with the $\alpha'$ corrections from the viewpoint of the world-sheet sigma model \cite{Lee:2019wij}.

With such an identification, one finds that at the infinite distance limit in the K\"ahler moduli space of $X_3$, characterized by the shrinking $K3$-fiber, i.e. $T_H \propto \text{Vol}(K3) \rightarrow 0$, the mass of the stringy state $M_{n=1}$ in units of 10d Type IIA string scale $M_s$ never vanishes as 
\be
\frac{M_{n=1}}{M_s} =\frac1 {RM_s}.
\ee

Hence one can conclude that the physical volume of the $K3$-fiber of $X_3$ never vanishes in the quantum moduli space if $R \neq \infty$. As the size of the M-theory circle $S^1$ grows infinity: $R\rightarrow \infty$, the 4d $N=2$ effective theory would regain one extra space and hence is in the disguise of the bona fide 5d $N=1$ theory, i.e. M-theory compactified on $X_3$. And at this limit, dubbed the"co-scaling" limit in \cite{Lee:2019wij}, quantum effects decouple and one recovers the classical result, i.e. the emergence of a tensionless heterotic string in M-theory. This is also echoed by an old observation stated in \cite{Witten:1996qb} that M-theory in 5d only "sees" the region at the infinity in the moduli space measured by the conformal field theory underlying the type IIA compactification.

Now one might tend to apply the same argument to a TCS $G_2$ manifold if a Type IIA orientifold limit exists. After all, we have argued from the previous sections that the infinite distance limit \eqref{infinitlimit} can be mimicked in the Type IIA orientifold setting, where the corresponding infinite distance limit is accompanied by the shrinking $K3$-fiber in the orientifold geometry. The fibre $K3$ on $X_3$ is even under the involution $\sigma$, and hence can be wrapped by D4-branes. One then expect to identify the mass of the particle from the wrapped D4-brane as the physical volume of the $K3$-fiber. Due to the same offset from the Casimir zero energy $E_0$, one would draw the same conclusion.

However, the above argument fails. Recall the uplifting of the cohomology groups from $X_3$ to $X$ in \eqref{hodgedecom}, one reads
\be
H^5(X)=H_-^4(X_3) \oplus H_-^1(S^1),
\ee
which implies that the emergent string from an M5-brane wrapping the $K3\times S^1$ on $X$ does not reduce to the particle from a D4-brane wrapping the corresponding $K3$-fiber in the orientifold $X_3/\sigma$, as the $K3$-fiber is even under the involution $\sigma$ and there is no any "even" one-cycle under the involution $-1$ in \eqref{typeIIAupliftg2} acting on the M-circle $S^1$. Instead, the emergent heterotic string in M-theory on $X$ reduces to the one from an NS5-brane wrapping the $K3$-fiber in Type IIA on $X_3/\sigma$ and one cannot use the same trick as the above.

Nevertheless, we can utilize string duality to solve the problem. Note that under the M/F-theory duality \eqref{stringduality}, the emergent tensionless heterotic string in the dual F-theory frame corresponds to a D3-brane wrapping $\mathbb P^1$ insider the base $B_3:=\mathbb P^1\times dP_9$ of the Donagi-Grassi-Witten manifold $X_4$. Hence we can turn the question to whether or not the volume $\text{Vol}(\mathbb P^1)$ receives quantum corrections at the infinite distance limit \eqref{infinitlimit2}. A previous study \cite{Mayr:1996sh} has been done for the cases of 4d non-critical strings in F-theory compactifications. In general, one can use mirror symmetry, now on the Calabi-Yau four-fold $X_4$, to do quantitative analysis of the quantum volume of $\mathbb P^1$. To this end, one generally needs to construct the mirror Calabi-Yau four-fold $\tilde{X_4}$ of $X_4$, and solve the periods from the Picard-Fuchs equations (possibly within the GKZ system \cite{Hosono:1994ax, Hosono:1993qy, Hosono:1995bm}) at the large complex structure regime and do the analytic continuation of the periods to the limit \eqref{infinitlimit2}, analogous to the calcuations of Calabi-Yau three-folds carried out in \cite{Klemm:2012sx, Corvilain:2018lgw, Lee:2019wij}. 

More straightforwardly, we instead use the same trick as the above to prove that the quantum volume of $\mathbb P^1$ does receive non-zero corrections. The basic idea goes like this: the particle from wrapping a D2-brane along the $\mathbb P^1$ on $X_4$ can be identified with the one from wrapping a D3-brane along $\mathbb P^1\times S_A^1$, with the help of the following duality  
$$
\label{FtypeIIAduality}
\text{F-theory on}~ X_4\times S_A^1\times S_B^1     \rightarrow \text{M-theory on}~ X_4\times S_B^1  \rightarrow  \text{Type IIA theory on}~X_4. 
$$
Then by the same token, we can determine the physical mass of the particle from the wrapped D2-brane by the mass formula in \eqref{massformular}.  Substituting the same Casimir zero energy $E_0=-1$ \footnote{One can also calculate the zero energy $E_0$ by going to dual F-theory picture, where the gravitational anomaly on the world-sheet (0,2) theory can be extrapolated from the detailed analysis in \cite{Weigand:2017gwb}, which is then given by \be 
E_0=-\frac{1}{2} C\cdot \bar{K}_{B_3}=-1, \ee where $C$ denote the class of the wrapped curve $\mathbb P^1$ and $\bar{K}_{B_3}$ stands for the anti-canonical divisor in the base $B_3:=\mathbb P^1\times dP_9$.} into \eqref{massformular}, the mass of the wrapped D2-brane with winding number $n=1$ is given by  
\be
\frac{M_{n=1}}{M_s} =\frac1 {R_A M_s},
\ee
where $R_A$ denotes the size of the one-cycle $S_A^1$. This implies that the mass of the wrapped D2-brane does not vanish (as $R_A \neq \infty$ in our case) at the limit \eqref{infinitlimit2}. Accordingly,  we conclude that the emergent heterotic string at the classical infinite distance limit  \eqref{infinitlimit}, identified with an M5-brane wrapping the shrinking $K3$-fiber on $X$, does receive quantum corrections and regain a non-zero tension at the quantum level \footnote{One may argue that at this limit \eqref{infinitlimit2}, the emergent heterotic string is not perturbative and hence the mass formula \eqref{massformular} does not apply to the limit. Indeed, as dubbed quasi-perturbative limit in \cite{Lee:2019tst}, certain non-perturbative effects such as the appearance of non-perturbative NS five-branes might occur.  However, as stated in \cite{Lee:2019tst}, one can be able to choose the extra parameters such that the heterotic string can be away from the five-branes and hence remains perturbative.}.  By the same argument, we further claim that all the tensionless heterotic solitonic strings in F-theory compactifications studied in \cite{Lee:2019tst}, which arise from wrapping D3-branes on shrinking two-cycles $C$ at the infinite distance limits with trivial normal bundles and $C\cdot \bar{K}_{B_3}=2$, receive quantum corrections and regain non-vanishing tensions at the quantum level.

Such quantum corrections enforce the total volume of $X$ to be divergent as now we have $\text{Vol}(X)\propto \text{Vol}(K3)\text{Vol}(S^3) \sim \infty$, leading to a decompactification (the gravity is decoupled in 4d). Indeed we can see this point from the scaling of the KK mass scale \eqref{KKtower}, where now it suffers further suppression from the infinite volume of $X$ and hence becomes the parametrically leading one for the infinite towers of the massless states. %Followed by this, an interesting question would be to explore what kind of field theory emerges at the limit \eqref{infinitlimit}, as gravity is decoupled.  

\subsection{Removing the Infinite Distance Limit}

 We have demonstrated that the quantum effects contribute a non-zero tension to the classical emergent tensionless heterotic string, and hence KK tower takes the role of the parametrically leading infinite tower of asymptotically massless states, which then signals a decompactification.  From a geometric perspective, the quantum corrections are expected to modify the metric of the classical moduli space such that the trajectories towards the limit \eqref{infinitlimit} are bent in some sense and hence the infinite limit is removed at the quantum level. This expectation does not come as a total surprise, as we have already seen several examples in a similar spirit. For example, Ooguri and Vafa \cite{Ooguri:1996me} proved that singularities at the conifold loci, which lie at the finite distance in the hypermultiplet moduli space $\mathcal M^Q$ in type IIA, can be smoothed by the sum of the E2-Instanton corrections, see also the mirror dual-type IIB case in \cite{Saueressig:2007dr}. More relevant, the smoothing of infinite distance singularities in the same hypermultiplet moduli space  $\mathcal M^Q$ by infinite (D(-1)/1-) instantons has also been verified nicely in \cite{Marchesano:2019ifh, Baume:2019sry}.  
 
In our cases, the K\"ahler  potentials $K^{G2}$ indeed receives quantum corrections from M2-brane instantons, which reduce to the perturbative $\alpha'$ corrections and the world-sheet instantons in a type IIA orientifold limit, and hence are expected to modify the metric in the moduli space $\mathcal M^{G2}$. The proof of such a smoothing of the infinite distance singularity is beyond the scope of this note. However, we would like to tentatively comment on the relevant parts in type IIA orientifolds. Recall that, before the orientifolding \eqref{orientifolding}, the Kahler potential $K$ in the K\"ahler moduli space $\mathcal M^K$ of a type IIA Calabi-Yau compactification reads
\be
K=-\text{ln}(i |Z^0|^2[2(\mathcal F-\bar{\mathcal F})-(\partial_A \mathcal F+\bar{\partial}_{\bar{A}}\bar{\mathcal F})(Z^A-\bar{Z}^A)]),\ee
where $\mathcal F$ denotes the N=2 prepotential, see the notations in the appendix \eqref{typeIIACY}. Away from the large volume regime in $\mathcal M ^K$, the perturbative $\alpha'$ corrections and non-perturbative $\mathcal F_{non}$ would dominate as the classical one $V_c:=-\frac16K_{ABC} t^A t^B t^C$ is suppressed by some small values of $t^A$. And indicated in \cite{Lee:2019wij }, the corrections modify the metric $g_{I\bar{J}}=\partial_I\bar{\partial}_{\bar J} \mathcal F_{non}$ such that they remove the singularities at the infinite distance characterized by a shrinking $K3$-fiber in $X_3$. 

Now in the 4d $N=1$ theories after the orientifolding \eqref{orientifolding}, quantum corrections are more drastic and complicated. Note that the K\"ahler potential $K^K$ of the truncated vector moduli space $\tilde{\mathcal M}^K$ typically receives various quantum corrections even at the large volume regime, as we briefly explain at the end of the appendix \ref{typeiiaorientifoldcompactific}. When approaching the large distance limit in $\tilde{\mathcal M}^K$ which mimics \eqref{infinitlimit}, one also expect that the descending quantum corrections from the N=2 theories modify the metric of $\tilde{\mathcal M}^K$ and remove the limit accordingly, though it is much more complicated to have a solid proof.

\section{Conclusion and Outlook}
\label{conclusionoutlook}
In this note, we have studied the Swampland Distance Conjecture, with the focus on testing the Emergent String Conjecture in TCS $G_2$ manifold compactifications of M-theory. We are interested in a (classical) infinite distance limit characterized by collapsing the (elliptic) $K3$-fiber of a $G_2$ manifold $X$, and we find a weakly coupled, tensionless fundamental heterotic string does emerge at the limit, whose light stringy excitations provide the parametrically leading one for infinite towers of asymptotically massless states, which is in line with the Emergent String Conjecture \cite{Lee:2019wij}. We further argue that such an emergent tensionless string receives quantum corrections and hence regains a non-zero tension by using M/F theory duality, which is in many ways similar to the 4d $N=2$ cases studied in \cite{Lee:2019wij}. As a by-product, we also claim all the tensionless heterotic critical strings arising from the wrapped D3-branes in F-theory compactifications regain non-zero tensions from quantum corrections at infinite distance limits by the same fashion. Such a result is also in line with the general expectation, that any BPS strings with $N=(0,2)$ supersymmetry in 4d $N=1$ settings, whether being critical or non-critical, do not receive protections from supersymmetry against quantum corrections, which are contrasted to their counterpart in 5d/6d supersymmetric theories. From a geometric perspective, the quantum corrections modify the K\"ahler potential $K^{G2}$ and are expect to remove the considered infinite distance limit at the quantum level. %We also comment on that in TCS $G_2$ compactifications, there could be another candidate of an infinite distance limit, characterized by the shrinking (putative) $T^4$-fibre, which might lead to an emergence of weakly coupled, tensionless type II fundamental string. 

Note that our infinite distance limit \eqref{infinitlimit} in a sense can be viewed as one inherited from a corresponding limit in the truncated K\"ahler moduli $\tilde{\mathcal M}^K$ in a type IIA orientifold limit, then one can wonder what happens to an infinite distance limit in the truncated hypermultiplet moduli space $\tilde{\mathcal M}^Q$ when uplifted to M-theory on a $G_2$ manifold. Such a large distance limit in type IIA orientifolds can be, for examples, obtained by asymptotically shrinking a non-contractible three-cycle, perhaps with the topology being $T^3$. If the three-cycle is odd under the orientifold involution $\sigma$, then we know from \eqref{hodgedecom}, it uplifts to a four-cycle on $G_2$ manifolds, with the topology possibly being $T^4$. And an M5-brane wrapping such a vanishing $coassoicative$ $T^4$-cycle would be a candidate for an emergent weakly coupled, tensionless fundamental Type II string \cite{Lee:2019wij}. Indeed, \cite{Braun:2017ryx, Braun:2017csz} have conjectured that the existence of a $T^4$-fibration by studying the mirror symmetry of TCS $G_2$ manifolds, which plays analogous roles as a special Lagrangian $T^3$-fibre in the SYZ conjecture \cite{Strominger:1996it}.  It would be very interesting to construct one from type IIA explicitly and study it in a similar vein.

We have ignored F-terms induced by both M2-brane instantons and background $G_4$ fluxes in this note.  In particular, a generic TCS $G_2$ compactification has been shown in \cite{Braun:2018fdp} that it receives infinite many M2-instantons corrections and the $E_8$ superpotential \cite{Donagi:1996yf} can be generated \footnote{More precisely, the prefactor of the $E_8$ superpotential would lose the universality for all M2-instanton contributions as it was expected in \cite{Donagi:1996yf}. The reason is that it receives extra different zero-modes from Ganor strings and as a result, the convergence to the $E_8$ superpotential breaks down. See more details in  \cite{Braun:2018fdp} }.  Presumably, such F-terms could also obstruct infinite distance limits in general. In 4d $N=1$ settings, recent work on this aspect has been carried out in \cite{Gonzalo:2018guu}, where they showed that the 4d $N=1$ scale potentials $V$ would be divergent at some infinite distance limits and claimed that the 4d dynamics would obstruct one from approaching any infinite distance limits. The main tool they have heavily utilized to lead to a divergent 4d scale potential $V$ is modular symmetry, under which the  potentials $V$ is invariant. The main contribution to the divergence of the 4d potential $V$ in \cite{Gonzalo:2018guu} comes from non-perturbative F-terms, whereas they have not considered quantum corrections to any 4d K\"ahler potentials yet. Nevertheless, this result does not conflict with our main result, as we conclude that in a sense, the quantum Kahler potential, including quantum corrections from M2-instantons, obstructs one from approaching infinite distance limits.  

Perhaps a more interesting question would rather be the study of effects induced by background $G_4$ fluxes on an emergent critical string in TCS $G_2$ compactifications. In F-theory Calabi-Yau four-fold compactifications, such a direction has already been studied in \cite{Lee:2019tst, Lee:2020gvu} with the main focus being the test of the WGC, where they have quantitatively analyzed the connections between modularities of the elliptic genera of an emergent fundamental heterotic string and enumerative invariants of a Calabi-Yau four-manifold. Without turning on background $G_4$ fluxes, the elliptic genus of an emergent heterotic string turns out to be trivial \footnote{ To see that, one recall that the elliptic genus of the emergent string in 4d F-theory compactification turns to be proportional to $\text{Tr} (Q)$, with $Q$ denoting charges under certain 4d (abelian) gauge groups \cite{Lee:2019tst, Lee:2020gvu}, which indicates the 4d effective theory is anomalous, and hence needs background $G_4$ fluxes to activate the 4d generalized Green-Schwarz mechanism \cite{Cvetic:2012xn} to cancel the gauge anomaly. If there is no background $G_4$ fluxes, then the elliptic genus is trivial.}.  It would be interesting to follow along this line in TCS $G_2$ compactifications and explore if there are any connections between elliptic genera of emergent strings and any "enumerative" invariants of TCS $G_2$ manifolds, for example, those uplifted from disk invariants in type IIA \cite{Mayr:2001xk, Aganagic:2001ug}. By switching on $G_4$ fluxes, it might be helpful to discuss stabilities for our infinite towers of massless states, which was studies in \cite{Grimm:2018ohb} as an implicit requirement for the SDC. We leave an investigation of them for the future.

\noindent {\bf Acknowledgements} 

We would like to thank Ling Lin and Timo Weigand for helpful collaborations at early stage of this project and beneficial discussions. We are indebted to Timo Weigand for sharing this topic. We would also like to acknowledge Daniel Klaewer and Max Wiesner for helpful conversations.  And we are grateful for nice hospitality from CERN theory division and the string group when this project was started.  
%\newpage
\appendix

\section{Type IIA Calabi-Yau Compactifications and Orientifolds }
\label{typeIIAorien}
\subsection{Type IIA Calabi-Yau Compactifications}
\label{typeIIACY}
In this appendix, we briefly summarize the relevant aspects of 4d $N=2$ moduli spaces of Type IIA Calabi-Yau compactifications and their reductions to 4d $N=1$ moduli spaces with orientifold compactifications. These topics have been extensively studied in the past and we would not list all the relevant references but instead follow the work \cite{Grimm:2004ua}.

The moduli space of 4d $N=2$ supergravity arising from a Calabi-Yau $X_3$ compactification of Type IIA has a local product structure
\be
\mathcal M^{N=2}=\mathcal M^K \times \mathcal M^Q,
\ee
where $\mathcal M^K$ is parametrized by the scalars in the vector multiplets of 4d $N=2$ supergravity and turns out to be special K\"ahler, and $\mathcal M^Q$ is spanned by the scalars in the hypermultiplets and is quaternionic. Notice that $\mathcal M^Q$ has a special K\"ahler submanifold which can also be viewed as $\mathcal M^K$ of the mirror manifold of $X_3$.  The SDC has been discussed, refined and verified firmly on the special K\"ahler moduli space $\mathcal M^K$ in \cite{Lee:2019wij, Corvilain:2018lgw, Grimm:2018cpv, Joshi:2019nzi} and on the quaternionic manifold $\mathcal M^Q$  \cite{Baume:2019sry, Marchesano:2019ifh}, respectively. 

The special K\"ahler manifold can be described by a single holomorphic function, dubbed prepotential $\mathcal F$. Equipped with a K\"ahler two-form $J$ and a complex structure three-form $\Omega^{3,0}$, the prepotential $\mathcal F$ at the large volume regime in the moduli space $\mathcal M^K$ has the following expression:  
\be
\mathcal F=-\frac16 K_{ABC} t^A t^B t^C+K_{AB}t^At^B+K_At^A+c+\mathcal F_{non},
\ee
where $K_{ABC}:= D_A\cdot D_B\cdot D_C$ denotes the triple intersection number between three divisors $D_A$ on $X_3$, $K_{AB}:=J\wedge D_A\wedge D_B$, $K_A:= \frac1{24}c_2(X_3)\cdot D_A$ with $c_2(X_3)$ denoting the second Chern class of $X_3$, and $c$ is a constant term $c:=i\frac{\zeta(3)}{(2\pi)^3}\chi(X_3)$. The last term $\mathcal F_{non}$ denotes non-perturbative contributions from the world-sheet instantons, which in general reads 
\be
\mathcal F_{non}=i\sum_{d_A \in H^2(X_3, \mathbb Z)} N_{\bf d}^0 \text{Li}(e^{\bf i d\cdot t}). 
\ee
Here $N_{\bf d}^0$  denotes the genus zero Gromov-Witten invariants of $X_3$, which can roughly be interpreted as enumerating rational curves $\mathbb P^1$ in the class $\bf d$. Note that at the large volume regime where $t^A \rightarrow i\infty$, such non-perturbative term is highly suppressed. The non-perturbative $\mathcal F_{non}$, however, can be calculated by mirror symmetry. 

For relevance, we also briefly mention the standard textbook-facts (see for example in \cite{Cox:2000vi}) on quantum volumes of 2n-cycles $C_{2n}$ on $X_3$, measured by the underlying world-sheet sigma model. They are not always measured by the K\"ahler form $J$ as classical volumes $\text{Vol}(C_{2n}): =\int_{C_{2n}} J^n$, but are only identical to $\text{Vol}(C_{2n})$ at the large volume regime. The classical volumes would be expected to receive significant corrections from quantum corrections such as the world-sheet instantons, which are not highly suppressed away from the large volume regime. One typically employ mirror symmetry to calculate such corrections.  Mirror symmetry states that the K\"ahler moduli space $\mathcal M^K (X_3)$ of Type IIA on $X_3$ is identical to the complex structure moduli space $\mathcal M^Q (\tilde{X_3})$ of Type IIB on the mirror manifold  $\tilde{X_3}$ of $X_3$. The important point is that the complex structure moduli space $\mathcal M^Q (\tilde{X_3})$ does not receives any quantum corrections, neither from the world-sheet instantons nor the space-time D-instantons. Therefore, it is classical and measured by the Calabi-Yau three-form $\tilde{\Omega}^{3,0}$ on $\tilde{X_3}$.  

One can choose an integral symplectic basis $\Gamma=(A^I, B_I) \in H_3(\tilde{X_3}, \mathbb Z)$ with the polarization 
\be
A^I \cap B_J=\delta^I_J.
\ee
Equipped with such a symplectic basis, one can define a set of natural coordinates on $\mathcal M^Q (\tilde{X_3})$ as
\be
Z^I:=\int_{A^I} \tilde{\Omega}^{3,0}, \qquad F_I:=\int_{B_I} \tilde{\Omega}^{3,0}, \qquad I=0,1,..., h^{2,1}(\tilde{X_3}).
\ee
Given such a set of coordinates one can then construct a period vector $\Pi=(Z^I, F_I)$ 
\be
\label{periodvector}
\Pi=\begin{pmatrix}
   Z^0       \\
     Z^A   \\
        F_A   \\
     F_0   \\
\end{pmatrix}, \qquad A=1,..., h^{1,2}(\tilde{X}{_3})=h^{1,1}(X_3),
\ee
which form a symplectic vector under $Sp(2(h^{1,2}(\tilde{X}{_3})+1), \mathbb Z)$. However, there is a redundancy associated with the Calabi-Yau three-form $\tilde{\Omega}^{3,0}$, as it is defined up to a complex rescaling. One can then use this rescaling to eliminant one of the periods $\Pi=(Z^I, F_I)$, say for example setting $Z^0=1$. Through this, one define a set of inhomogeneous, flat coordinates $t^A=\frac{Z^A}{Z_0}, A=1,..., h^{1,2}(\tilde{X_3})$, which characterizes complex structure deformation of $\tilde{X_3})$.  According to mirror symmetry,  the large complex structure limit of $\mathcal M^Q (\tilde{X_3})$ maps to the large volume limit in $\mathcal M^K (X_3)$. And at this large limit, the above flat coordinates coincident with the classical, complexified K\"ahler parameters $t^A$ in $\mathcal M^K(X_3)$ from the reduction of $J_C=iJ+B_2$ along two-cycles on $X_3$. 

Now under the mirror symmetry, the component of this period vector $\Pi$ can measure the quantum volumes of the $2n$-cycles ($n=0,1,2,3$) on $X_3$, which is a set of solutions of a Picard-Fuchs equation and can be solved by various methods.  Away from the large complex structure limit, one can use the analytic continuations of the periods to define the coordinates over the full quantum K\"ahler moduli space  $\mathcal M^Q (\tilde{X_3})$, see more details with some explicit examples of elliptically fibered Calabi-Yau three-folds studied in \cite{Klemm:2012sx, Lee:2019wij}. 

The Kahler potential $K$ on $\mathcal M^K$ is then determined by 
\be
K=-\text{ln}(i 2(\mathcal F-\bar{\mathcal F})-(\partial_A \mathcal F+\bar{\partial}_{\bar{A}}\bar{\mathcal F})(Z^A-\bar{Z}^A)).
\ee
At the large volume regime, the K\"ahler potential $K$ reduces to the classical one:
\be
K=-\text{ln}(-\frac16\int_{X_3} J\wedge J\wedge J).
\ee

\subsection{Type IIA Orientifold Compactifications}
\label{typeiiaorientifoldcompactific}
In this subsection, we review some relevant basic aspects of type IIA orientifolds for our $G_2$ compactifications. The orientifold compactifications can be obtained from Calabi-Yau compactifications by modding out a orientifolding $\mathcal O$, which is given by 
\be
\label{orientifolding}
\mathcal O=\Omega_p(-1)^{F_L}\sigma,
\ee
where $\Omega_p$ denotes the world-sheet parity, $(-1)^{F_L}$ is the fermion number operator on the left-moving sector and $\sigma$ denotes an involution on $X_3$. In order to preserve 4d $N=1$ supersymmetry, the involution $\sigma$ acts on the K\"ahler form $J$ and the Calabi-Yau three-form $\Omega^{3,0}$ of $X_3$ as follows
\be
\sigma^\ast J=-J, \qquad  \sigma^\ast \Omega^{3,0}= e^{2i\theta}\bar{\Omega}^{3,0},
\ee
where $\theta$ is a certain constant phase. Under such involution $\sigma$, the cohomology (homology) groups of $X_3$ split into even and odd eigen-spaces as 
\be
H^p(X_3)=H_+^p(X_3)\oplus H_-^p(X_3), \qquad p=2, 3.
\ee

Such the anti-holomorphic involution $\sigma$ generically leads to special Lagrangian submanifolds $\Lambda_3$ of $X_3$ which home the fix points of $\sigma$, and satisfy 
\be
J|_{\Lambda_3}=0, \qquad \text{Im}(e^{i\theta} \Omega^{3,0})|_{\Lambda_3}=0,
\ee
and O6-planes/D6-branes wrap such special Lagrangian cycles $\Lambda_3$ .  

The involution $\sigma$ also acts non-trivially on various form fields in 10d type IIA supergravity as
\be
\begin{split}
\sigma: &B_2 \rightarrow -B_2, C_1 \rightarrow -C_1,  \\
&C_3 \rightarrow C_3,  g \rightarrow g, \phi \rightarrow \phi. 
\end{split}
\ee
Especially, under the involution $\sigma$, we have a decomposition for the complexified K\"ahler form $J_c:= B_2+iJ_2$ as 
\be
J_c=(b^a+i v^a)\omega_a:=t^a\omega_a, \qquad a=1,..., h^{1,1}_-(X_3).
\ee
As results of the fact that the volume $J\wedge J\wedge J$ being odd under the involution $\sigma$ and the hodge duality, one can infer that $h^{1,1}_-=h^{2,2}_+$,  $h^{1,1}_+=h^{2,2}_-$ and $h^{2,1}_-=h^{2,1}_+=h^{2,1}+1$.

 After the orientifolding \eqref{orientifolding}, the moduli space $ \mathcal M^{N=1}$ of a corresponding 4d $N=1$ theory can be viewed as the truncation of $\mathcal M^{N=2}$ and also has the local product structure 
 \be
 \mathcal M^{N=1}=\tilde{\mathcal M}^K \times \tilde{\mathcal M}^Q,
 \ee
 where $\tilde{\mathcal M}^K $ is a subspace of the $\mathcal M^K $ with dimension $h^{1,1}_-(X_3)$, which is trivially truncated and hence remains as a special K\"ahler manifold. The K\"ahler potential of $\tilde{\mathcal M}^K $ at the large volume regime is given by 
 \be
 K^K=-\text{ln}[-\frac43\int_{X_3}(J\wedge J\wedge J)]=-\text{ln}[\frac i6 \mathcal K_{abc}(t-\bar t)^a(t-\bar t)^b(t-\bar t)^c].
 \ee
 Whereas $\tilde{\mathcal M}^Q$ is a subspace of the quaternionic manifold $\mathcal M^Q$ with dimension $h^{1,2}(X_3)+1$, whose K\"ahler potential $K^Q$ at the large volume regime is given by 
 \be
K^Q=-2\text{ln}(2\int_{X_3}\text{Re}(C\Omega^{3,0})\wedge \ast \text{Re}(C\Omega^{3,0}))=-\text{ln}e^{-4D},
\ee
where $C$ is a complex compensator which offsets redundancies of $\Omega^{3,0}$ in order to have the above closed form and $D$ denotes the 4d dilaton. 

Obviously, the overall K\"ahler potential $K^{\text{IIA}}$ in the type IIA orientifold is given by the sum of the above two as 
\be
K^{\text{IIA}}=K^K+K^Q=-\text{ln}(-\frac43\int_{X_3}(J\wedge J\wedge J))-2\text{ln}(2\int_{X_3}\text{Re}(C\Omega)\wedge \ast \text{Re}(C\Omega)).\ee

We are interested in quantum corrections to the K\"ahler potential $K^K$ in $\tilde{\mathcal M^K}$. Similar to $K$ in the 4d $N=2$ moduli space $\mathcal M^K$, it is expected to receive contributions from world-sheet instanton and perturbative $\alpha'$ corrections, whereas D2-brane instantons correct $K^Q$ as they couple to three-cycles, measured by $\Omega^{3,0}$. Note that even at the large volume regime of $\tilde{\mathcal M^K}$, the K\"ahler potential $K^K$ would receive non-perturbative corrections which are not highly suppressed in contrast with the one in $K$ prior to the orientifolding \eqref{orientifolding}. To see that, note that at the large volume regime of $\tilde{\mathcal M^K}$, not all $v^A$ would be large as one have $v^\alpha=0, \alpha \in H_{2}^+(X_3)$ in the orientifold $X_3/\sigma$, hence some quantum corrections at $t^a=-b^a$ are not necessarily suppressed anymore, which is in contrasted with the ones in $\mathcal M^K$. For example, the prepotental $\mathcal F$ has the following surviving terms at the large volume regime
\be
\mathcal F_{non}=\sum_{\beta \in H_2^+(X_3, \mathbb Z)} n_{\beta}^0 \text{Li}_3(e^{i k_\alpha t^\alpha}),
\ee 
where $k_{\alpha}=\int_{\beta} \omega_\alpha$. Away from the large volume regime, one can expect more drastic corrections. However, the explicit calculations are more complicated than ones in the $N=2$ settings.

%%%%%%%%%%%%%%%%%%%%%%%%%%%%%%%%%%%%%%%%%%%%%%%
%%%%%%%%%%%%%%%%%%%%%%%%%%%%%%%%%%%%%%%%%%%%%%%
%%%%%%%%%%%%%%%%%%%%%%%%%%%%%%%%%%%%%%%%%%%%%%%
%\newpage
\clearpage
\phantomsection
\addcontentsline{toc}{section}{References}
\bibliography{References}  
\bibliographystyle{custom1}

\end{document}